\documentclass[prb,preprint]{revtex4-1} 

\usepackage{amsmath,amsthm,amssymb,amsfonts}
\usepackage{mathrsfs}
\usepackage{graphicx}
\usepackage{xcolor}
\usepackage{setspace}
\usepackage{fancyhdr}
\usepackage{hyperref}
\usepackage{subfig}
\usepackage{adjustbox}
\DeclareSymbolFontAlphabet{\mathrsfs}{rsfs}
\DeclareMathAlphabet{\mathcal}{OMS}{cmsy}{m}{n}
\newcommand{\scri}{\mathrsfs{I}}

\newcommand{\be}{\begin{equation}}
\newcommand{\ee}{\end{equation}}

\begin{document}
\title{Hyperbolic times in Minkowski space}
\author{An\i l Zengino\u{g}lu}

\email{anil@umd.edu} 
\affiliation{Institute for Physical Science and Technology, University of Maryland, College Park, MD 20742, USA}
\date{\today}


\begin{abstract}
Time functions with asymptotically hyperbolic geometry play an increasingly important role in many areas of relativity, from computing black-hole perturbations to analyzing wave equations.
Despite their significance, many of their properties remain underexplored.
In this expository article, I discuss hyperbolic time functions by considering the hyperbola as the relativistic analog of a circle in two-dimensional Minkowski space and argue that suitably defined hyperboloidal coordinates are as natural in Lorentzian manifolds as spherical coordinates are in Riemannian manifolds. 

\end{abstract}

\maketitle

\section{Hyperbolic geometry in relativity}
The discovery of hyperbolic geometry is one of the most impactful developments in the history of mathematics. Its revelation by Gauss, Bolyai, and Lobachevsky in the 19th century was the culmination of a story spanning over 2,000 years that rivals any multi-generational science fiction saga in drama and excitement \cite{rosenfeld2012history}. The subsequent analysis of non-Euclidean geometry by Beltrami, Poincaré, and Riemann was the necessary mathematical development that prepared the ground for special and general relativity as physical theories of space, time, and gravity. 

Even though Minkowski space is flat, hyperbolic geometry plays an important role in special relativity that Minkowski was keenly aware of \cite{gray_non-euclidean_1999, galison2010minkowski}. For example, the interpretation of Lorentz transformations as hyperbolic rotations gives an intuitive demonstration of relativistic phenomena such as velocity addition or Thomas precession \cite{rhodes_relativistic_2004, ungar2008analytic, dray_geometry_2017, Moore2022}. Hyperbolic geometry not only provides the kinematic space of special relativity \cite{gray_non-euclidean_1999}, it also serves as a model for space that connects a source of radiation to an idealized observer (see Sec.~\ref{sec:snapshot}). 

General time functions that share the asymptotic properties of hyperbolas are called hyperboloidal \cite{friedrich1983cauchy}. The behavior of such hypersurfaces makes them appealing to describe global aspects of spacetimes, including asymptotic boundaries, black hole horizons, and cosmological horizons. In suitable hyperboloidal coordinates, space flows toward null horizons similarly as in the river model of black holes \cite{martel2001regular,hamilton2008river}. Hyperboloidal time functions play an increasingly important role in relativity across many active research areas such as black-hole perturbation theory, gravitational waves, and the mathematical analysis of wave equations \cite{gautam2021summation, macedo2022hyperboloidal, jaramillo2022pseudospectrum, peterson20233d, markakis2023symmetric, panosso2024hyperboloidal, lefloch2023euclidean, vano2023conformal, kroon2024v}. 

In this expository article, I present various aspects of hyperbolic time functions in the simple case of two-dimensional Minkowski space at a level suitable for an advanced course on special relativity. The guiding principle is that the hyperbola in Minkowski space is the analog of a circle in Euclidean space. The main objectives are to demonstrate the essential properties of such surfaces and to clarify misconceptions. Most of the discussion requires no more than a basic understanding of Minkowski space and coordinate transformations.

Our discussion starts in Sec.~\ref{sec:standard} with a quick introduction to Minkowski space and its compactification using Penrose coordinates—an essential tool for understanding the global causal structure of spacetimes. We consider the standard coordinates and their weaknesses in the context of the global causal structure of Minkowski space: the standard time coordinate in Minkowski space is degenerate at infinity. Time functions that respect time translation symmetry must approach null infinity to resolve the asymptotic coordinate singularity. 
In Sec.~\ref{sec:hyperbola}, we introduce spacelike hyperbolas as analogs of circles in Euclidean space. We clarify their global causal properties (\ref{sec:spacelike}), demonstrate their necessity for far-away observers (\ref{sec:snapshot}), and present a variational principle underlying their construction as solutions to the isoperimetric problem in Minkowski space (\ref{sec:maximal}). 
Section \ref{sec:time_functions} compares Milne slicing (\ref{sec:slicing}) and hyperbolic foliation (\ref{sec:foliation}), revealing that the analogy between the circle and the hyperbola is misleading when constructing smooth time functions describing the asymptotic domain. The impact of the asymptotic coordinate singularity in Milne slicing is demonstrated on a simple wave equation in (\ref{sec:energy}. Global energy is conserved in standard time and Milne slicing but decays due to radiation to infinity in hyperbolic foliation. In Section \ref{sec:penrose_coordinates}, we revisit the construction of Penrose coordinates and recognize that they arise from a combination of Milne slicing and hyperbolic foliation. The paper ends with conclusions highlighting the importance of hyperboloidal time functions in relativity (\ref{sec:conclusions}).

\section{The standard time coordinate in Minkowski space}\label{sec:standard}
\begin{figure}
        \includegraphics[width=0.45\textwidth]{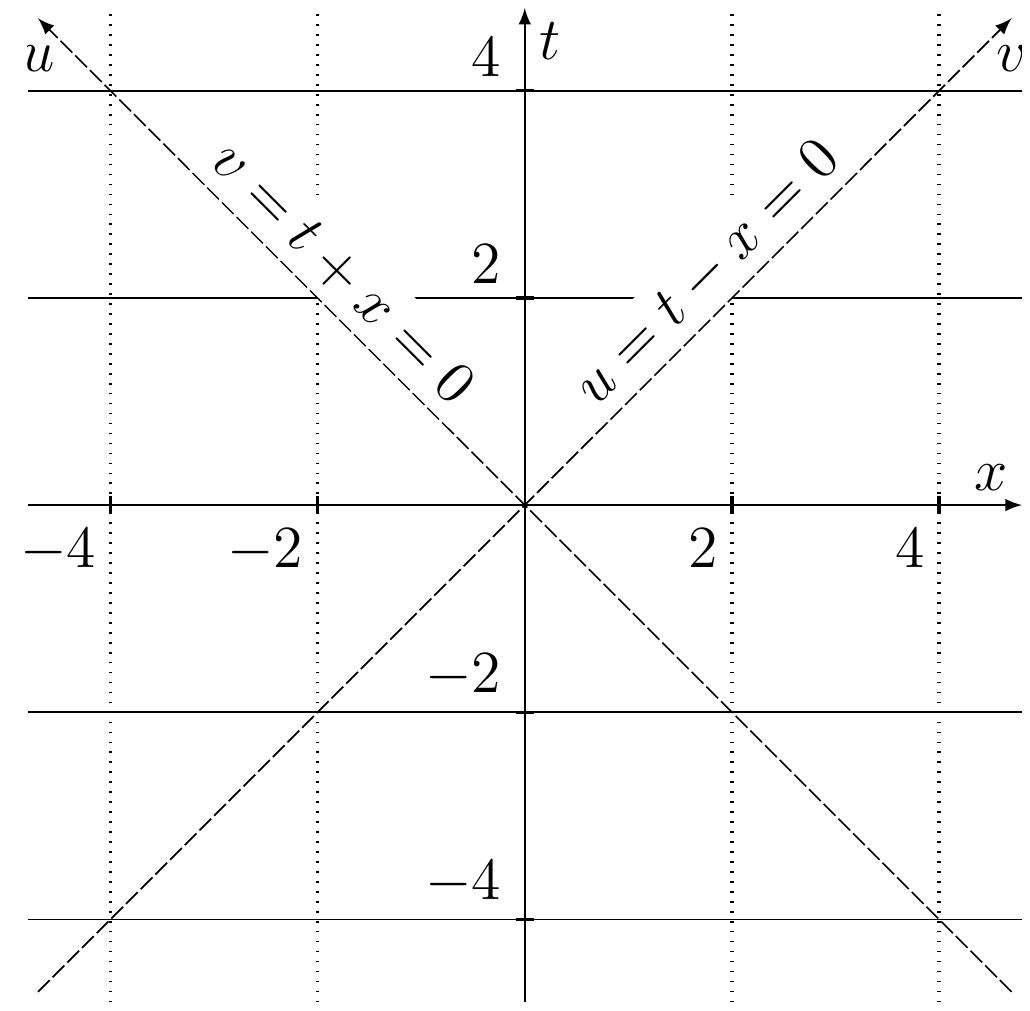}
    \hspace{7mm}
        \includegraphics[width=0.45\textwidth]{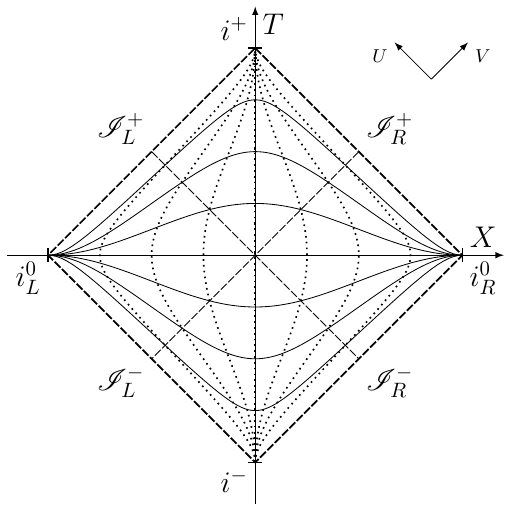}
    \caption{The left panel shows the standard Minkowski coordinate grid in $\{t,x\}$ and the null cone through the origin representing the axes of null coordinates $\{u,v\}$. The right panel shows the level sets of standard coordinates in a Penrose diagram $\{T, X\}$. In all figures, solid, horizontal curves are spacelike; dotted, vertical curves are timelike; and dashed curves are null. Slices of the standard time coordinate $t$ intersect at spatial infinities, suggesting that the time coordinate $t$ is unsuitable for the global structure.}
    \label{fig:standard}
\end{figure}

Our stage is the two-dimensional Minkowski space. The Minkowski metric in standard coordinates $t\in(-\infty,\infty)$ and $x\in(-\infty, \infty)$ is\footnote{We use natural units in which the speed of light, $c$, and the gravitational constant,$G$, are set to unity.}
\be\label{eqn:minkowski} ds^2 = -dt^2 + dx^2. \ee
The Minkowski metric differs from the Euclidean metric by the presence of a time direction. As a consequence, the metric has a null space consisting of null rays: $v=t+x$, $u=t-x$ (Fig.~\ref{fig:standard}). The null space plays a fundamental role in Minkowski space with no equivalent structure in the Euclidean case. In particular, the analog of a circle in Minkowski space becomes an unbounded curve (Sec.~\ref{sec:spacelike}). Therefore, it is essential to understand the global structure of spacetime.

A common way to represent the global causal structure of Minkowski space is to draw a Penrose diagram: a conformal diagram using Penrose coordinates $\{T,X\}$. The key idea is conformal compactification, which maps the infinite Minkowski space to a finite space with a boundary representing infinity \cite{penrose_asymptotic_1963, penrose2011republication}. Many such mappings are available, but 
the historical choice by Penrose is the tangent function applied to the null coordinates. Specifically, we introduce coordinates $U=\tan^{-1} u$ and $V=\tan^{-1} v$ which map the infinite interval $(-\infty,\infty)$ to the finite, open interval $(-\pi/2,\pi/2)$. Squeezing infinite distances to a finite range leads to a singular metric at the limiting points
\be\label{eqn:penrose_metric} ds^2 = -du \, dv = -\left(\frac{1}{\cos U \, \cos V}\right)^2 dU \, dV.\ee
We capture this singular behavior in a coordinate-dependent scale factor called the conformal factor $\Omega = \cos U \, \cos V$. Considering the conformally rescaled, regular metric $\bar{ds}^2 = \Omega^2 ds^2$, we add the limiting points of the open interval to our domain and obtain the compact interval $[-\pi/2,\pi/2]$, completing the conformal compactification.

One typically introduces time and space coordinates, $T=U+V$ and $X=V-U$, with the range $|T|+|X| \in[0,\pi]$, to draw the Penrose diagram (right panel of Fig.~\ref{fig:standard}). The conformal boundary is the zero set of the conformal factor given by $|U| = |V| = \pi/2$, or $|T|+|X|=\pi$.  
This boundary is not part of Minkowski space but represents the asymptotic limit of its geodesics. Spacelike geodesics end at spatial infinity, $i^0_{\{R,L\}} = \{T=0, X=\pm \pi\}$; timelike geodesics end at timelike infinity, $i^\pm = \{T=\pm \pi, X=0\}$; and null geodesics end at null infinity denoted by $\scri^\pm_{\{R,L\}}$. Null infinity is also called scri for the ``script I" symbol $\scri$ used to denote it.

A big advantage of conformal compactification is that it replaces asymptotic limits such as $x\to\infty$ or $u\to\infty$ with local analysis at the boundary. Penrose coordinates perform this compactification along the null coordinates so that null rays are straight lines at 45 degrees.
The Penrose diagram of Minkowski space is discussed in many textbooks on relativity \cite{wald1984general, blau2011lecture,carroll2019spacetime}. An advanced discussion of conformal methods in general relativity can be found in the monograph \cite{kroon2017conformal}.

Penrose coordinates for time $T$ and space $X$ are not direct compactifications of the standard coordinates $t$ and $x$. There is an intermediary step involving null directions. It is instructive to see why direct compactification does not work. Setting $t=\tan \bar{t}$ and $x=\tan \bar{x}$, we get
\[ ds^2 = - \frac{d\bar{t}^2}{\cos^4 \bar{t}} + \frac{d\bar{x}^2}{\cos^4 \bar{x}}.  \]
One cannot capture the singularity of this metric at the boundary $|\bar{t}|=|\bar{x}|=\pi/2$ in a conformal factor. The underlying geometric reason is that the standard coordinates are singular at their asymptotic endpoints, $i^0_{R,L}$ and $i^\pm$, where their level sets intersect. The intersection of the level sets can be seen on the right panel of Fig.~\ref{fig:standard}. This problem affects any effort to construct a time foliation that asymptotes to spatial infinity and respects time symmetry. We discuss the case of spatial infinity below as we are primarily interested in time functions, but similar arguments apply to the timelike infinities, $i^\pm$. In Sec.~\ref{sec:penrose_coordinates}, we will see that Penrose coordinates are direct compactifications of hyperboloidal coordinates.

\vspace{3mm}

\noindent \textbf{The standard time coordinate is singular at spatial infinity.}\\
Level sets of $t$ intersect at the spatial infinities, $i^0_{\{R,L\}}$, as seen on the Penrose diagram in Fig.~\ref{fig:standard}. This intersection is reminiscent of radial rays in polar coordinates and indicates that the time coordinate $t$ is unsuitable for the global causal structure. 

Looking at the left panel of Fig.~\ref{fig:standard}, students find it hard to believe that the $t$ slices are singular at infinity because they seem parallel and non-intersecting on the plot. However, any coordinate system will look like a Cartesian grid when plotted with respect to itself. 
We can see the singular behavior only when we compare the coordinates to another, more suitable system. This is true for standard coordinates of Minkowski space as it is for polar coordinates of Euclidean space. 

The reason for the coordinate singularity at spatial infinity is the vanishing of the generator of time symmetry, $\partial_t$. Expressing it in Penrose coordinates, we get
\be\label{eqn:delt}\partial_t = \cos^2 U \, \partial_U + \cos^2 V \, \partial_V = \left(1+\cos(2T)\cos(2X)\right)\partial_T - \sin(2T)\sin(2X)\partial_X. \ee
All components of $\partial_t$ vanish at spatial and timelike infinities indicating that these points are fixed points of the time symmetry causing the intersection of time slices. A regular time coordinate that approaches spatial infinity cannot preserve the time-translation symmetry.


\vspace{3mm}
\noindent \textbf{The standard time coordinate is not operationally meaningful for far-away observers.}\\
The level sets of $t$ represent space for globally synchronized inertial observers at rest relative to each other. This notion of ``now'' as a snapshot in time has been abstracted from our everyday experience with weak gravitational fields, slow speeds, and small distances. Its construction requires the synchronization of clocks based on mirrors and is, therefore, operationally meaningful only in the vicinity of an inertial observer. However, as an observer moves farther from a source, one cannot construct such a global time function operationally. For example, astrophysical sources of gravitational radiation are typically thousands or millions of light years away, so synchronization of clocks with such sources is not feasible. Asymptotic observers have only access to the null rays that reach them from the source.

In the next sections, we will see that spacelike hyperbolas resolve these problems.

\section{Spacelike hyperbola}\label{sec:hyperbola}

\begin{figure}
    \centering
        \includegraphics[width=0.45\textwidth]{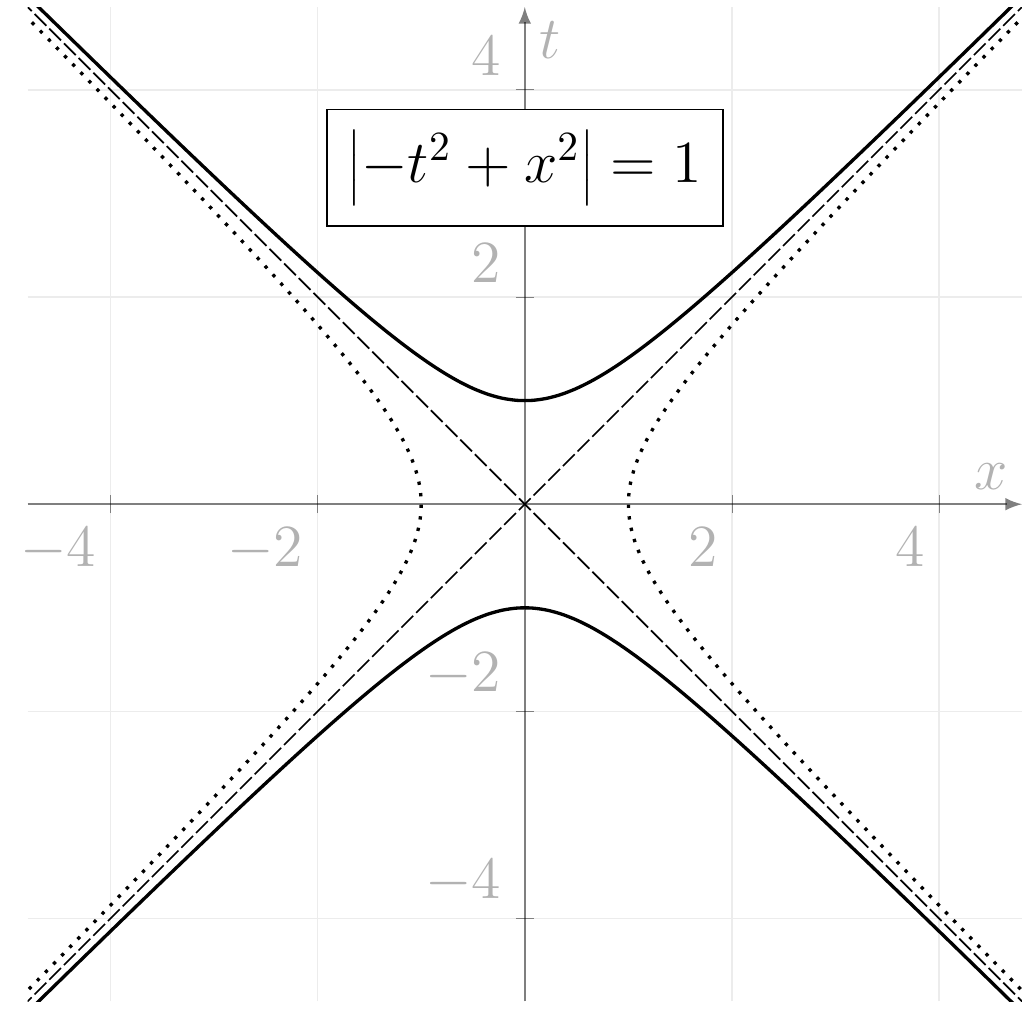}
    \hspace{7mm}
        \includegraphics[width=0.45\textwidth]{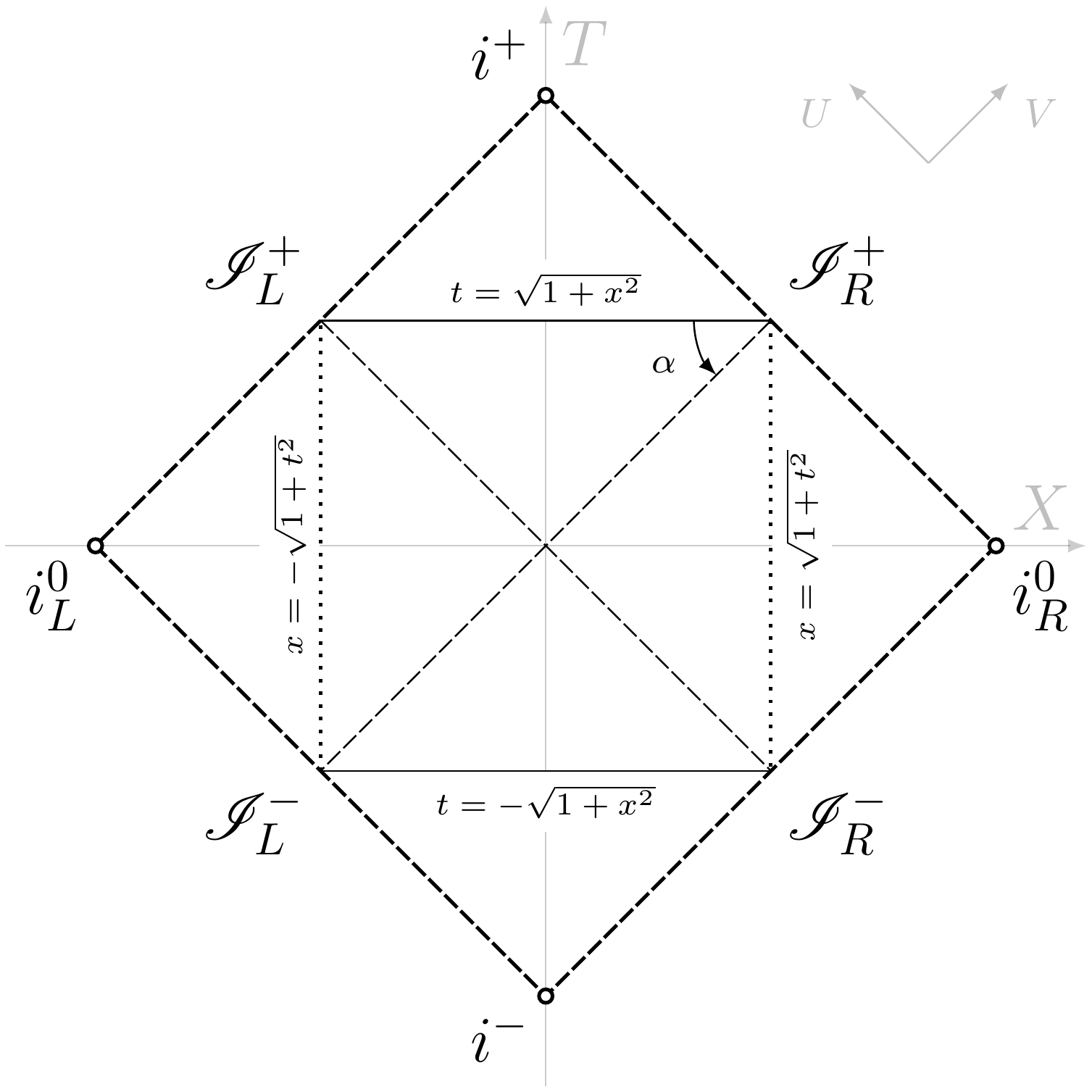}
    \caption{The unit hyperbola $\left|-t^2 + x^2\right| = 1$ consists of four curves. The solid curves inside the null cone, $|t|>|x|$, are spacelike. The dotted curves outside the null cone, $|x|>|t|$, are timelike. 
    The unit hyperbola---the analog of the unit circle in Euclidean space--- is a square on the Penrose diagram of Minkowski space. The future spacelike hyperbola has the angle $\alpha = -\pi/4$ to the $V$-axis at future null infinity. The diagram demonstrates that hyperbolas do not change their causal structure asymptotically. Their causal properties extend to the asymptotic domain.}
    \label{fig:unit_hyperbola}
\end{figure}

In a historical talk during the eightieth meeting of the Assembly of Natural Scientists and Physicians in Cologne in 1908, Minkowski presented the unification of space and time into spacetime \cite{gray_non-euclidean_1999, galison2010minkowski}. He demonstrated the invariance of his new metric under Lorentz transformations using the unit hyperbola $t^2 - x^2 = 1$. The hyperbola plays a fundamental role in Minkowski space as the set of points equidistant from a fixed point, representing the analog of a circle in Euclidean space \cite{Moore2022}. It is invariant under Lorentz transformations, just as a circle is under rotations. In contrast to the circle, it is an unbounded curve with negative curvature and two asymptotes, $t\pm x=0$, which are the null rays emanating from the origin. The equation for the hyperbola with (pseudo-)radius $\eta$ centered at the origin is 
\be \label{eqn:hyperbola} \left|t^2 - x^2\right| = \eta^2. \ee
The unit hyperbola with $\eta=1$ is depicted in Fig.~\ref{fig:unit_hyperbola} in standard coordinates on the left and Penrose coordinates on the right. Its asymptotes are plotted as dashed lines emanating from the origin. It is neat that the analog of the unit circle in Euclidean space is a square on the Penrose diagram of Minkowski space. 

We are primarily interested in spacelike hyperbolas for constructing time functions. The unit hyperbolas centered at the origin are spacelike in the interior of the null cone with $|t|>|x|$. Future hyperbolas are in $t>|x|$, and past hyperbolas are in  $t<-|x|$  (see the shaded regions in Fig.~\ref{fig:hyperbolic_slicing}). The following sections discuss three properties of the future hyperbola relevant for time functions for idealized, asymptotic observers: everywhere spacelike (\ref{sec:spacelike}), a snapshot in time (\ref{sec:snapshot}), and a maximal curve satisfying a variational principle (\ref{sec:maximal}). These properties suggest that hyperboloidal coordinates are as natural in Lorentzian manifolds as spherical coordinates are in Riemannian manifolds.

\subsection{Spacelike hyperbolas are globally spacelike}\label{sec:spacelike}
The future hyperbola on the left panel of Fig.~\ref{fig:unit_hyperbola} seems to become ``asymptotically null.'' The curve clearly asymptotes to the null cone and, therefore, to null infinity. In that sense, one could call it asymptotically null \cite{calabrese2006asymptotically, misner2006excising}. However, it is important to emphasize that this is not a causal statement. The perception that the hyperbola becomes causally null in the asymptotic domain arises from its representation in standard coordinates. The global causal nature of the hyperbola cannot be discussed faithfully in standard coordinates as they are unsuitable in the asymptotic domain. It is evident from the Penrose diagram on the right panel of Fig.~\ref{fig:unit_hyperbola} that the future hyperbola is spacelike everywhere, including the asymptotic domain, because the curve representing the hyperbola intersects null infinity horizontally. In contrast, an asymptotically null curve has a null tangent space and is diagonal at null infinity.

We can explicitly calculate the angle at which the hyperbola transverses null infinity \cite{zenginouglu2007conformal, kroon2017conformal}. The calculation is easiest in compactified null coordinates $\{U, V\}$. Spacelike hyperbola of radius $\eta$ satisfy 
\[ \eta^2 = t^2  - x^2 = u\, v = \tan U \, \tan V. \]
We calculate the angle of incidence $\alpha$ that the future hyperbola makes with the $V$-axis at the right future null infinity, $\scri^+_R=\{V=\pi/2\}$ (see Fig.~\ref{fig:unit_hyperbola}). An asymptotically null curve would have an angle of $0$ or $-\pi/2$. The graph $U(V)$ for constant values of $\eta^2$ has an angle of incidence given by
\[ \tan \alpha|_{\scri_R^+} = \frac{dU}{dV}\Bigg|_{V=\frac{\pi}{2}}  = \frac{d}{dV} \left[\tan^{-1} \left(\frac{\eta^2}{\tan V}\right) \right] \Bigg|_{V=\frac{\pi}{2}} = - \eta^2. \]
For any non-vanishing, finite $\eta$, this angle is in the range $(-\pi/2,0)$. For the unit hyperbola with $\eta=1$, we have $\alpha = -\pi/4$, giving the horizontal line in the Penrose diagram Fig.~\ref{fig:unit_hyperbola}. At the limiting values of $\eta$ we get asymptotically null curves: for $\eta=0$ the curve becomes tangent to the $V$-axis with $\alpha=0$; for $\eta\to\infty$, the curve becomes tangent to the $U$-axis with $\alpha\to -\pi/2$. Curves for different values of $\eta$ are depicted in Fig.~\ref{fig:hyperbolic_slicing}, where one can see this behavior. Note that the particular value of the angle is not important for the qualitative discussion. The important point is that the future hyperbola is spacelike everywhere, including in the asymptotic domain.

Asymptotically null curves in the Penrose compactification have a vanishing $V$-derivative at future null infinity, implying that the linear term in the Taylor series of the graph $U(V)$ around $V=\pi/2$ vanishes. We translate this local condition in compactified coordinates to an asymptotic condition in standard coordinates to give a condition for an embedded curve, $t(x)$, that behaves like the future hyperbola. For such a curve to be spacelike at null infinity, we require that the $1/x$-term does not vanish in the expansion of the embedding,
\be\label{eqn:t_expansion} t(x) = x + \frac{C}{x} + \mathcal{O}(x^{-2}), \quad C \neq 0, \qquad \textrm{as} \quad x\to\pm\infty. \ee
The constant term is absorbed in the definition of the time function. As an example, below is the large $x$ expansion for the unit hyperbola
\[ t(x) = \sqrt{1+x^2} = x + \frac{1}{2 x} + \mathcal{O}(x^{-3}). \]
The non-vanishing $1/x$ term ensures that the curve is spacelike at null infinity. 

One can understand that the hyperbola does not become asymptotically null by considering hyperbolic geometry. The hyperbola, or its higher-dimensional version, the hyperboloid, represents a model for hyperbolic geometry—a \emph{homogeneous} space of constant negative curvature \cite{rosenfeld2012history}. The induced metric on the hyperboloid is the Riemannian metric of hyperbolic space, indicating that the hyperboloid is a spacelike surface everywhere, including the ideal points at infinity. This property of the hyperboloid allows us to use it as a snapshot in time.

\subsection{The future hyperbola is a natural snapshot in time}\label{sec:snapshot}
A distinction between space and time for an idealized observer along null infinity necessarily involves a spacelike curve extending from null infinity toward the source of radiation. We have seen that hyperbolas inside the null cone are spacelike curves everywhere. Any spacelike curve can serve as a snapshot, but generally, the associated observers are not naturally related to each other. The hyperbola has a particular property that provides a natural notion of a snapshot in Minkowski space. 

\begin{figure}
    \centering
    \begin{adjustbox}{raise=0.34cm}
        \includegraphics[width=0.45\textwidth]{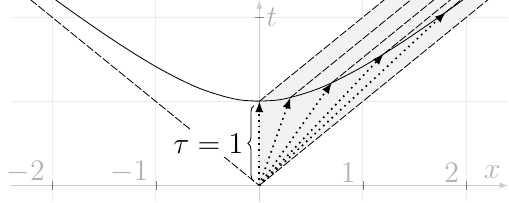}\hspace{7mm}
    \end{adjustbox}
        \includegraphics[width=0.45\textwidth]{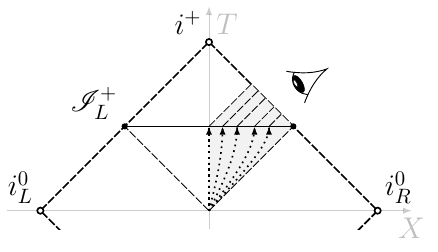}
        \caption{Timekeepers synchronize their clocks at the origin and travel towards an observer at different relative speeds. They send light signals to the idealized observer at infinity at their unit proper time. The spacelike curve that connects the timekeepers is the unit hyperbola, which provides a natural snapshot for idealized observers at infinity.}
    \label{fig:proper}
\end{figure}

Consider a family of observers with synchronized clocks at the origin, called timekeepers. The timekeepers move toward an idealized observer to the right at different speeds. These speeds can be stochastically distributed. Each timekeeper sends a light signal to the idealized observer at infinity at their proper time $\eta$. The observer receives these signals at different times within the gray area in Fig.~\ref{fig:proper}. The curve that connects the timekeepers at their equal proper time and serves as a snapshot for the observer is the future hyperbola
$ \eta = \sqrt{t^2 - x^2}$ (see Fig.~\ref{fig:proper}).
We can pick any other point $\{t_0,x_0\}$ to synchronize the timekeepers and perform a similar construction with $ \eta = \sqrt{(t-t_0)^2 - (x-x_0)^2}$.

This notion of a snapshot works well with asymptotic observers of radiation. Any time function constructed by an observer at infinity is related to a hyperbolic time function by a smooth transformation. In contrast, the transformation from the standard time coordinate to hyperbolic or null coordinates is singular at infinity. Therefore, a time function for observers at infinity must resemble a hyperbola in its asymptotic behavior.


\subsection{The hyperbola is a maximal curve}\label{sec:maximal}

The circle is the simplest nontrivial example of a minimal curve solving the isoperimetric problem in the plane: enclosing the largest area with a fixed perimeter \cite{blaasjo2005isoperimetric}. Since the hyperbola is the relativistic analog of a circle, we may expect that the hyperbola solves a similar type of isoperimetric problem. 

We can think of the isoperimetric problem as a constrained variational problem. An equivalent problem to maximizing the area with a fixed perimeter is minimizing the perimeter with a fixed area. The action for a minimal curve with length $L$ enclosing a fixed area $A$ is
\be\label{eqn:action} S = L - \lambda \cdot A.  \ee
The solution of this minimization problem is a circle of radius $1/\lambda$. The inverse radius, or equivalently, the Lagrange multiplier of the minimization problem $\lambda$, is the constant curvature of the circle.

In Minkowski space, minimizing length is not an interesting problem because null rays have zero length. Instead, in Lorentzian variational problems, we look for a maximizing curve. The action for maximizing boundary length is the same as \eqref{eqn:action}, except that the length $L$ and the area $A$ are computed for the Minkowski metric \eqref{eqn:minkowski}. This idea generalizes to higher dimensions and curved spacetimes, leading to the analysis of constant mean curvature surfaces in general relativity \cite{brill_k_1980, malec2009general}.

\begin{figure}
    \centering
    \begin{adjustbox}{raise=0.2cm}
    \includegraphics[width=0.4\textwidth]{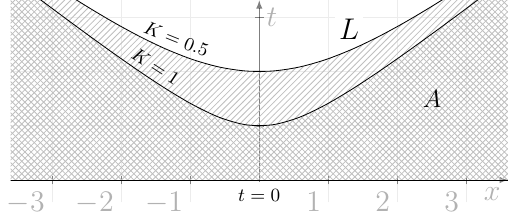}
    \end{adjustbox}\hspace{7mm}
    \includegraphics[width=0.5\textwidth]{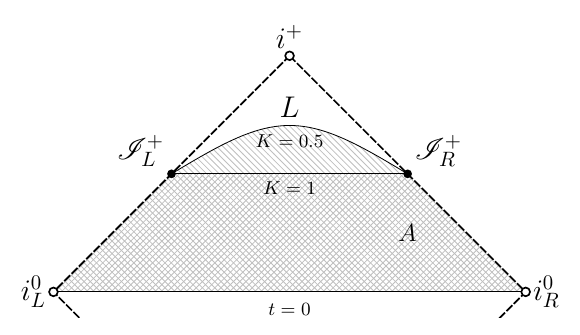}
    \caption{The isoperimetric problem in two-dimensional Minkowski space is to maximize the length $L$ of a curve $t(x)$ that encloses a given spacetime area $A$ with respect to a reference surface, arbitrarily taken as the $t=0$ slice. Hyperbolas are solutions to this isoperimetric problem and have constant mean curvature. The upper hyperbola with a smaller mean curvature $K=0.5$ encloses a larger spacetime area than the lower hyperbola with a larger mean curvature $K=1$.}
    \label{fig:soap}
\end{figure}

Consider a curve parametrized by $x$ as $t=t(x)$. The length of the curve is
\[ L = \int ds = \int \sqrt{-dt^2 + dx^2} = \int_{-\infty}^\infty \sqrt{-t'(x)^2+1} \ dx, \] 
where $t'(x) = dt(x)/dx$. We need a reference curve to calculate the enclosed spacetime area. Since we are performing the calculation in standard coordinates, we take the $t=0$ line as a reference (see Fig.~\ref{fig:soap}), but the outcome is independent of the reference curve. The enclosed spacetime area by the curve $t(x)$ and the line $t=0$ is 
\be\label{eqn:area} A = \int_{-\infty}^\infty \, dx \int_{0}^{t(x)} dt = \int_{-\infty}^{\infty} t(x) dx. \ee
We can write the action \eqref{eqn:action} as an integral over a Lagrangian density functional
$$ S = \int_{-\infty}^\infty \mathscr{L}\, dx = \int_{-\infty}^\infty \left[\sqrt{1-t(x)'^2} - \lambda t(x)\right] \, dx. $$
The Euler-Lagrange equation for varying $t(x)$ with respect to the parameter $x$ reads
$$ \frac{d}{dx} \frac{\partial\mathscr{L}}{\partial t'} = \frac{\partial \mathscr{L}}{\partial t} \quad \Rightarrow \quad
 \frac{d}{dx} \frac{t'}{\sqrt{1-t'^2}} = \lambda. $$
We integrate by $x$ and solve for $t'$ with the boundary condition $t'(0)=0$
$$ t'(x) = \pm \frac{\lambda x}{\sqrt{1+ \lambda^2 x^2}}. $$
Choosing the positive sign and integrating again, we obtain the future hyperbola 
\be\label{eqn:embedded} t(x) = \sqrt{\frac{1}{\lambda^2} + x^2}. \ee

The Lagrangian multiplier $\lambda$ is inversely related to the radius of the hyperbola $\eta$ and represents the constant mean extrinsic curvature of the curve, $K=\lambda$, just as in the Euclidean case. The variational problem \eqref{eqn:action} does not determine the Lagrangian multiplier, $\lambda$, or equivalently, the mean curvature $K$. Its value depends inversely on the spacetime area $A$ we prescribe in the constraint\footnote{We can compute the dependence of spacetime area $A$ on mean curvature $K$ by evaluating the integral \eqref{eqn:area} using the curve \eqref{eqn:embedded} and taking out the leading term. The result is $\mathcal{O}(\log K/K^2)$}. The larger the spacetime area to be enclosed, the smaller the mean curvature of the hyperbola. In Fig.~\ref{fig:soap}, the upper line with $K=0.5$ encloses a larger area than the lower line with $K=1$. The maximal area is enclosed by a curve with vanishing mean curvature, $K=0$. Such maximal slices are level sets of $t$.

\section{Hyperbolic time functions}\label{sec:time_functions}
The future hyperbola is spacelike everywhere and has a natural interpretation as a snapshot in time constructed by an idealized observer at infinity. We want to describe dynamical evolution by parametrizing spacetime using a foliation of such snapshots as given by level sets of a time function. The time function should be suitable for describing the global causal structure in an operationally meaningful way. We have seen that the standard time coordinate is unsuitable for this purpose. We now construct time functions based on hyperbolas and study a simple example of an evolution equation based on them. At the end of the section, we recognize the nature of Penrose coordinates as hyperboloidal.

\subsection{Milne slicing}\label{sec:slicing}

\begin{figure}
    \centering
    \begin{adjustbox}{raise=0.6cm}
    \includegraphics[width=0.4\textwidth]{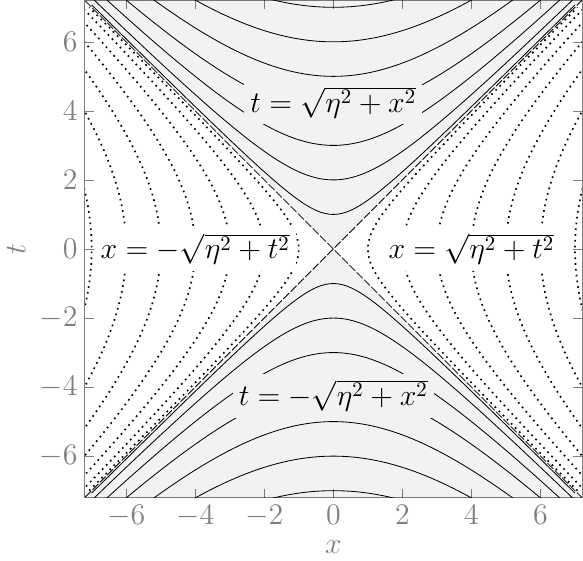}
    \end{adjustbox}\hspace{1cm}
    \includegraphics[width=0.5\textwidth]{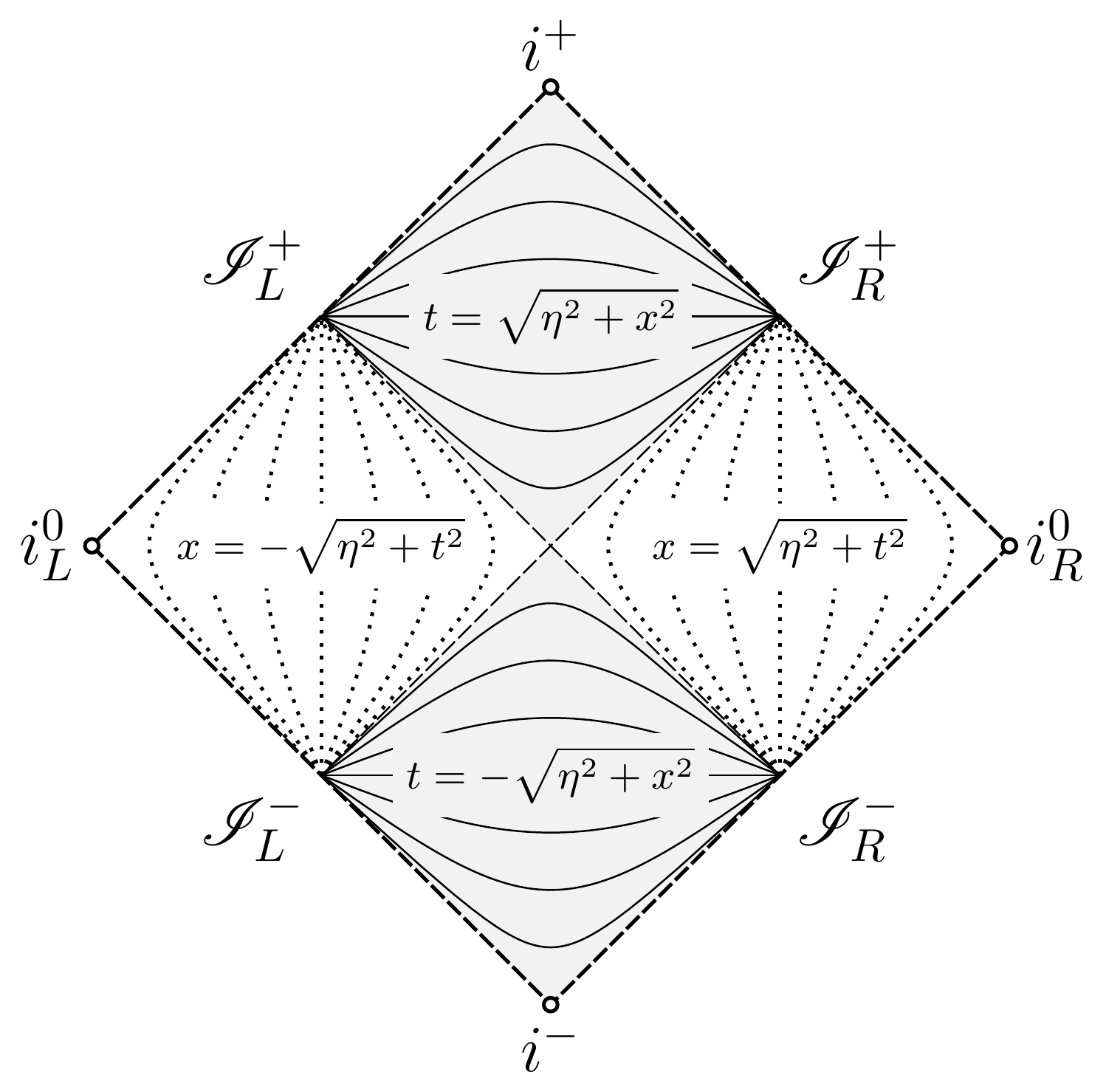}
    \caption{Slicing of Minkowski space by hyperbola of different radii (Milne slicing). The slices asymptote to the same null cone, intersecting at null infinity and becoming singular. The slicing does not respect the time-translation symmetry of Minkowski space, and the metric depends on Milne time \eqref{eqn:milne}.}
    \label{fig:hyperbolic_slicing}
\end{figure}

The first attempt to construct a suitable time function based on hyperbolas is to parametrize Minkowski space by hyperbolas of different radii, as in the case of Euclidean space with polar coordinates based on circles of different radii. This approach goes back to Milne \cite{milne1936relativity}, who considered the resulting metric a cosmological model \cite{possel2019teaching} \footnote{Many early works on hyperboloidal methods used a similar approach \cite{dirac1949forms, chen_hyperboloidal_1971, strichartz_harmonic_1973}. In higher dimensions, Milne slicing leads to a metric whose time slices have the same geometry as Anti-de Sitter time slices. Therefore, this approach is appealing in recent works on quantum field theory and flat-space holography \cite{de_boer_holographic_2003, cheung_4d_2017, ogawa_wedge_2023}.}

The hyperbola of radius $\eta$ is given by the equation \eqref{eqn:hyperbola}. The corresponding parametrization of Minkowski space is depicted in Fig.~\ref{fig:hyperbolic_slicing}. The figure shows that Milne slices have different causal properties in different parts of the light cone centered at the origin. Outside the light cone, Milne slices are timelike curves representing uniformly accelerated observers. These domains are known as Rindler wedges \cite{rindler1960hyperbolic}. The light cone acts as a null horizon for the Rindler wedges.

Inside the light cone, Milne slices are spacelike curves and provide a time function. We are particularly interested in parametrizing the future light cone by hyperbolas with Milne time $\eta=\sqrt{t^2-x^2}$. Parametrizing the future light cone by the proper time of observers with arbitrary velocity seems a natural choice (compare Sec.~\ref{sec:snapshot}). Each slice has a constant mean extrinsic curvature, and the value of the curvature depends on time as $K=1/\eta$. 

To write the Minkowski metric in Milne slicing in its simplest form, we define the comoving coordinate $\chi$ via $t=\eta \cosh\chi$ and $x = \eta \sinh\chi$. The metric becomes
\be\label{eqn:milne} ds^2 = -d\eta^2 + \eta^2 d\chi^2. \ee

Milne slicing has some undesirable properties for describing time evolution. First, we see from Fig.~\ref{fig:hyperbolic_slicing} that the slices intersect at null infinity and do not provide a smooth foliation of the conformal boundary. This behavior of Milne slices near null infinity is similar to standard time slices near spatial infinity. The evolution vector field, $\partial_\eta$, vanishes at null infinity, indicating that the slices intersect there.

The second problem is that Milne slicing does not preserve the time-translation symmetry, leading to a time-dependent metric \eqref{eqn:milne}. While the choice of coordinates is a matter of taste and depends on the problem, introducing an artificial time dependence into the metric is undesirable for many applications. 

A related problem is that the light cone at the origin plays a unique role in Milne slicing. This is similar to the role that the origin plays in polar coordinates, around which rotational symmetry is defined. However, it is unclear why the origin should play such a unique role when we describe time evolution in Minkowski space. Note that polar coordinates parametrize all of space, whereas Milne time is restricted to inside the light cone. In the following section, we will resolve these issues by constructing a hyperbolic foliation that respects the time translation symmetry of Minkowski space and is regular at null infinity.
 
\subsection{Hyperbolic foliation}\label{sec:foliation}
The Milne model \eqref{eqn:milne} uses the radii of hyperbolas as the time coordinate, similar to how polar coordinates use the radii of circles as a space coordinate. However, Euclidean space has no null cone or a time direction. The time-translation symmetry in Minkowski space plays a central role in many applications, and we should keep it in our coordinates.

\begin{figure}
    \centering
    \begin{adjustbox}{raise=0.8cm}
    \includegraphics[width=0.4\textwidth]{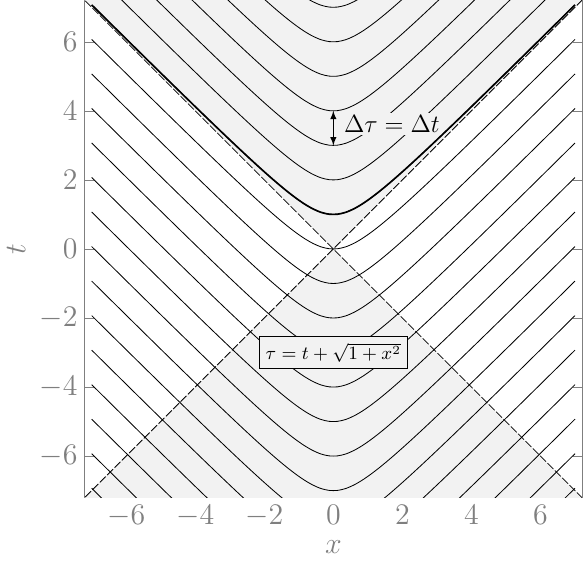}
    \end{adjustbox}\hspace{1cm}
    \includegraphics[width=0.5\textwidth]{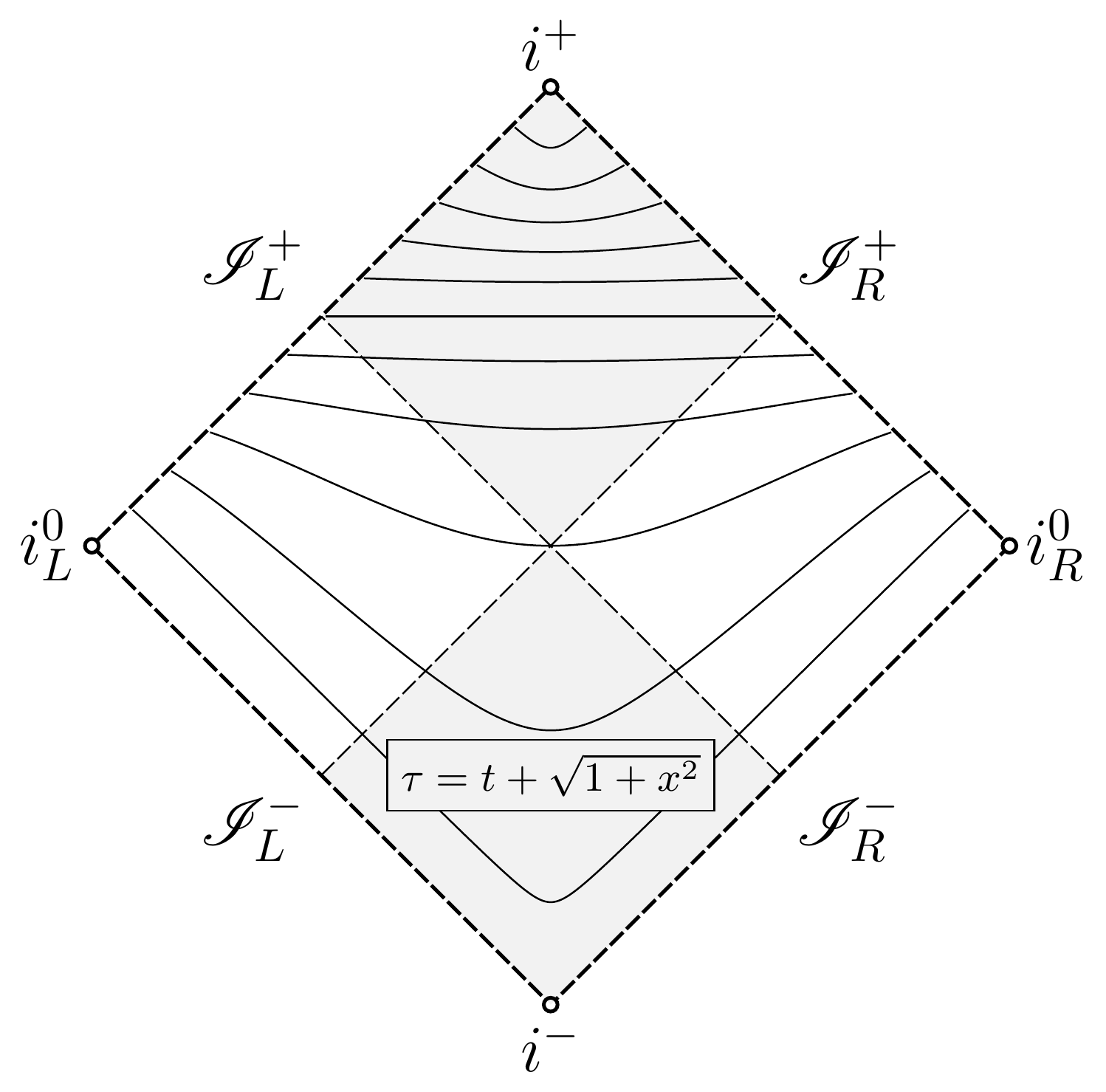}
    \caption{Hyperbolic foliation \eqref{eqn:shifted} fixes a future hyperbola and drags it along the time direction $\partial_t$. The resulting metric \eqref{eqn:hyperboloidal} is independent of time. The null cone through the origin is shaded in gray to contrast with the Milne slicing of Fig.~\ref{fig:hyperbolic_slicing}.}
    \label{fig:hyperboloidal_foliation}
\end{figure}

Instead of slicing the spacetime by hyperbolas of different radii, we fix the radius of a hyperbola to some value, $L$, and shift it along the time direction by $\tau$ \cite{gowdy_wave_1981},
\be\label{eqn:shifted} (t-\tau)^2 - x^2 = L^2. \ee
We call the slicing of Minkowski space with the time function $\tau$ a hyperbolic foliation. Such time-shifted hyperbolas can be generalized to higher dimensions and curved, stationary spacetimes, leading to hyperboloidal foliations \cite{zenginouglu2008hyperboloidal}. Solving the above equation for $\tau$, we can choose the sign for future or past hyperbolic foliations that foliate future or past null infinity. The future hyperbolic foliation is depicted in Fig.~\ref{fig:hyperboloidal_foliation}. Each slice asymptotes to a future null cone shifted in time. We write the time function $\tau$ as
\be\label{eqn:hyperbolic_foliation} \tau = t  - \sqrt{L^2+x^2} = t - \frac{L}{\cos\rho}, \ee
where $x = L \tan \rho$ is a scaled spatial compactification. 

This time function has many appealing properties. It preserves the generator of the time-translation symmetry, $\partial_t = \partial_\tau$. The mean extrinsic curvature is constant in space and time, $K=1/L$. The metric is time-independent by construction
\begin{align}\label{eqn:hyperboloidal} 
    ds^2 &= - d\tau^2 - \frac{2x}{\sqrt{L^2 + x^2}} \, d\tau\, dx + \frac{L^2}{L^2 + x^2} \, dx^2 \\
    &= \frac{1}{\cos^2 \rho} \left( -\cos^2 \rho \, d\tau^2 -2 L \sin\rho\, d\tau d\rho + L^2 d\rho^2  \right). \end{align}
We can absorb the radius of the hyperbola, $L$, in the conformal factor using a rescaled time coordinate $\bar\tau = L\tau$. As we have seen in Sec.~\ref{sec:snapshot}, the radius of the hyperbola corresponds to the proper time from the origin, so the conformal metric with the rescaled time is scale-free. The conformal metric is regular at null infinity and includes a non-diagonal term indicating that space flows outward in the future hyperbolic foliation. This outflow behavior is analogous to the inflow behavior in the river model of black holes where regular foliations respecting time symmetry have a diagonal shift term \cite{martel2001regular, hamilton2008river}.

\subsection{Energy conservation and decay}\label{sec:energy}
Noether's theorem states that the symmetries of the background imply conserved quantities for the dynamics. In Minkowski space, we have a time-translation symmetry implying conservation of energy. It is a standard result that energy is conserved for non-dissipative systems. As a simple demonstration, consider the homogeneous wave equation for a massless scalar field $\phi(t,x)$
\[ -\phi_{tt} + \phi_{xx} = 0, \]
where the subscript is shorthand for a partial derivative. This equation was first written by d'Alembert in 1747 to describe the problem of a vibrating string, long before the unification of space and time into spacetime \cite{oliveira2020d}. Remarkably, the scalar wave equation, with its finite propagation speed, is inherently relativistic in natural units. We write the equation as a conservation law to demonstrate energy conservation
\[ \mathcal{E}_t = \mathcal{F}_x, \qquad \textrm{with} \qquad \mathcal{E} = \frac{1}{2} \left(\phi_t^2+\phi_x^2\right), \quad \mathcal{F} = \phi_t \phi_x.\]
This form of the wave equation gives us the energy density $\mathcal{E}$ and the flux density $\mathcal{F}$.\footnote{We can also compute these quantities from the energy-momentum tensor of a scalar field, but the formulation as a conservation law seems simpler requiring only the formation of total derivatives in the wave equation \cite{bizon_saddle-point_2010}.} The total energy is the integral of the energy density over space, $E = \int_{-\infty}^{\infty}\mathcal{E}\, dx$. The scalar field in the standard time coordinate has globally conserved energy, assuming that the boundary terms vanish due to the finite speed of propagation,
\[ \frac{dE}{dt} = \int_{-\infty}^\infty \frac{d \mathcal{E}}{dt}\, dx = \int_{-\infty}^\infty \frac{d\mathcal{F}}{dx}\, dx = \phi_t \phi_x \Big|_{-\infty}^\infty = 0. \]
However, our experience with isolated dynamical systems suggests they lose energy to radiation. We want a formula that relates the energy decay to radiation flux at null infinity. In null coordinates, the energy calculated along null infinity (also called Bondi energy after \cite{bondi1962gravitational}) readily captures such decay. We have seen that the standard time coordinate does not capture the asymptotic behavior adequately. One might anticipate, therefore, that slices approaching null infinity instead of spatial infinity will capture the energy loss. To see if that is the case, we compute the wave equation in Milne coordinates
\[ - \phi_{\eta\eta} + \frac{1}{\eta^2}\phi_{\chi\chi} - \frac{1}{\eta} \phi_\eta = 0. \]
We compactify space with $\sinh\chi = \tan \rho$ to analyze the asymptotic behavior, 
\[ - \frac{\eta^2}{\cos\rho}\phi_{\eta\eta} + \cos\rho\, \phi_{\rho\rho} - \frac{\eta}{\cos\rho}\,\phi_\eta - \sin\rho\, \phi_{\rho}= 0. \]
The energy and flux densities are
\[ \mathcal{E} = \frac{1}{2}\left( \frac{\eta^2}{\cos\rho} \phi_\eta^2 + \cos\rho\, \phi_\rho^2 \right), \qquad \mathcal{F} = \cos\rho\, \phi_\eta \phi_\rho. \]
Energy is conserved also in Milne time,
\[ \frac{dE}{d\eta}  =  \int_{-\frac{\pi}{2}}^{\frac{\pi}{2}} \frac{d\mathcal{F}}{d\rho} \ d\rho =  \cos \rho\ \phi_\eta \phi_\rho \Big|_{-\frac{\pi}{2}}^{\frac{\pi}{2}}  = 0. \]

Energy conservation is not just a consequence of evaluating the energy at spatial infinity. It is a consequence of the intersection of time slices in the asymptotic domain. To model energy radiation to infinity, we need an asymptotically regular foliation. Switching to a hyperbolic foliation makes the loss of energy explicit. 
The transformed wave equation in hyperbolic foliation $\tau$ and compactifying coordinate $\rho$ reads after a division by $\cos^2\rho$
\[ -\phi_{\tau\tau} - 2 \sin\rho\, \phi_{\tau\rho} + \cos^2\rho \,\phi_{\rho\rho} - \cos\rho \,\phi_\tau - 2\sin\rho \cos\rho\, \phi_\rho \phi = 0.\]
The energy and flux densities are
\[ \mathcal{E} = \frac{1}{2} \left( \phi_\tau^2 + \cos^2 \rho \,\phi_\rho^2\right), \qquad \mathcal{F} = - \sin\rho\,\phi_\tau^2 + \cos^2\rho\, \phi_\tau \phi_\rho\,. \]
Now we can show that a Bondi-type total energy is not conserved but decays due to radiation across null infinity,
\[ \frac{d}{d\tau} E  =  \int_{-\frac{\pi}{2}}^{\frac{\pi}{2}} \frac{d\mathcal{F}}{d\rho} \ d\rho = \mathcal{F} \Big|_{-\frac{\pi}{2}}^{\frac{\pi}{2}}  = - \phi_\tau^2\Big|_{\scri_R^+} - \phi_\tau^2\Big|_{\scri_L^+} \leq 0. \]

\begin{figure}
    \includegraphics[width=\textwidth]{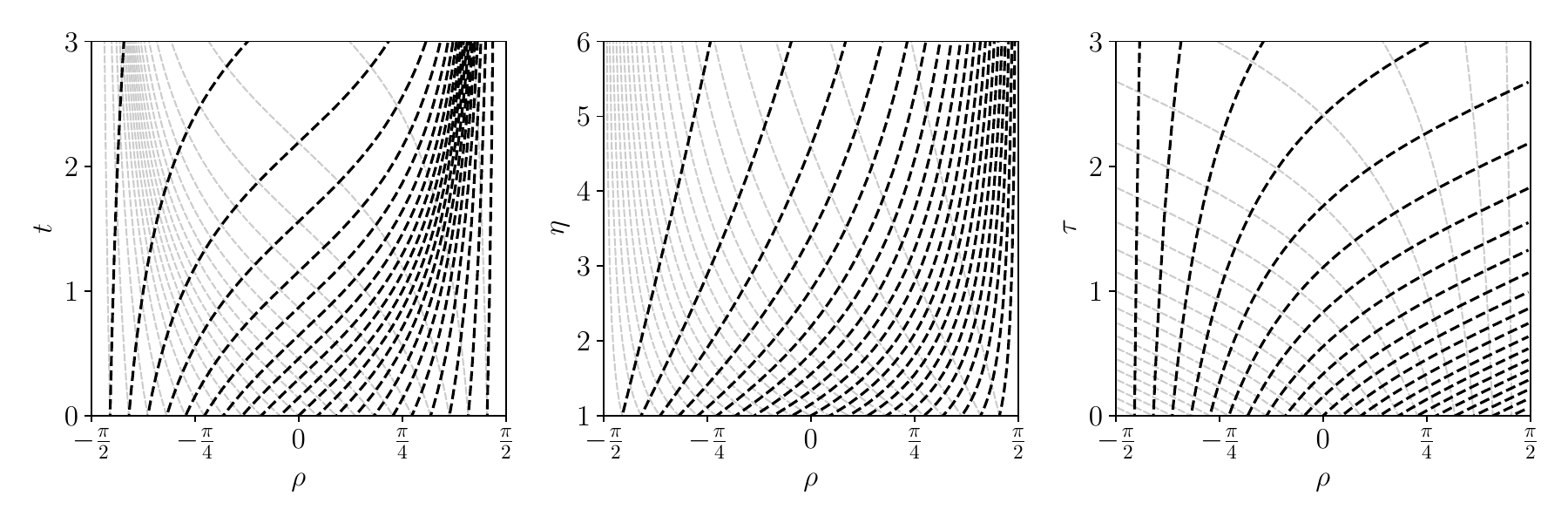}
    \caption{Null rays in the compactified space coordinate $\rho\in(-\pi/2,\pi/2)$ for the standard time $t$ on the left, Milne slicing $\eta$ in the middle, and hyperbolic foliation $\tau$ on the right. Outgoing null rays in standard time and Milne slicing never reach the conformal boundary. Compare to the Penrose diagrams in Figs.~\ref{fig:standard}, \ref{fig:hyperbolic_slicing}, and \ref{fig:hyperboloidal_foliation}.}    \label{fig:characteristics}
\end{figure}

One can understand the behavior of energy by visually inspecting the Penrose diagrams in Figs.~\ref{fig:standard}, \ref{fig:hyperbolic_slicing}, and \ref{fig:hyperboloidal_foliation}. An outgoing characteristic in standard time and Milne time crosses all slices because each level set of the corresponding time function ends at the same asymptotic point. In hyperbolic foliation, however, an outgoing null ray carrying energy will intersect only a finite number of slices. Therefore, energy will decay along a hyperbolic foliation. The propagation of such energy packages along null rays is best visualized by the characteristics of the metric plotted in Fig.~\ref{fig:characteristics}. Outgoing characteristics leave the domain through the conformal boundary only in the case of the hyperbolic foliation (right-most panel).

\subsection{Penrose coordinates are hyperboloidal}\label{sec:penrose_coordinates}

Penrose diagrams are important tools to visualize the global causal structure of spacetimes. The associated coordinates are constructed through an intermediate step of null coordinates \cite{penrose_asymptotic_1963,penrose_zero_1965} as reviewed in Sec.~\ref{sec:standard}. We saw in that section that Penrose coordinates do not arise from the compactification of standard coordinates. What is then the nature of the coordinates that lead to Penrose coordinates when compactified? 

\begin{figure}
    \begin{adjustbox}{raise=0.6cm}
        \includegraphics[width=0.4\textwidth]{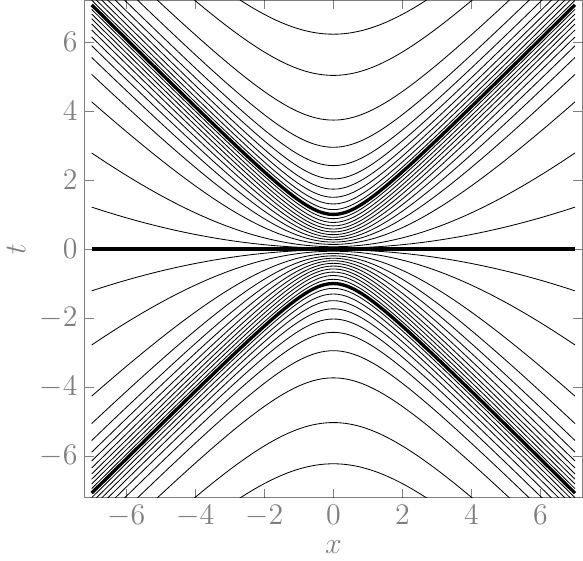}
    \end{adjustbox}\hspace{1cm}
    \includegraphics[width=0.5\textwidth]{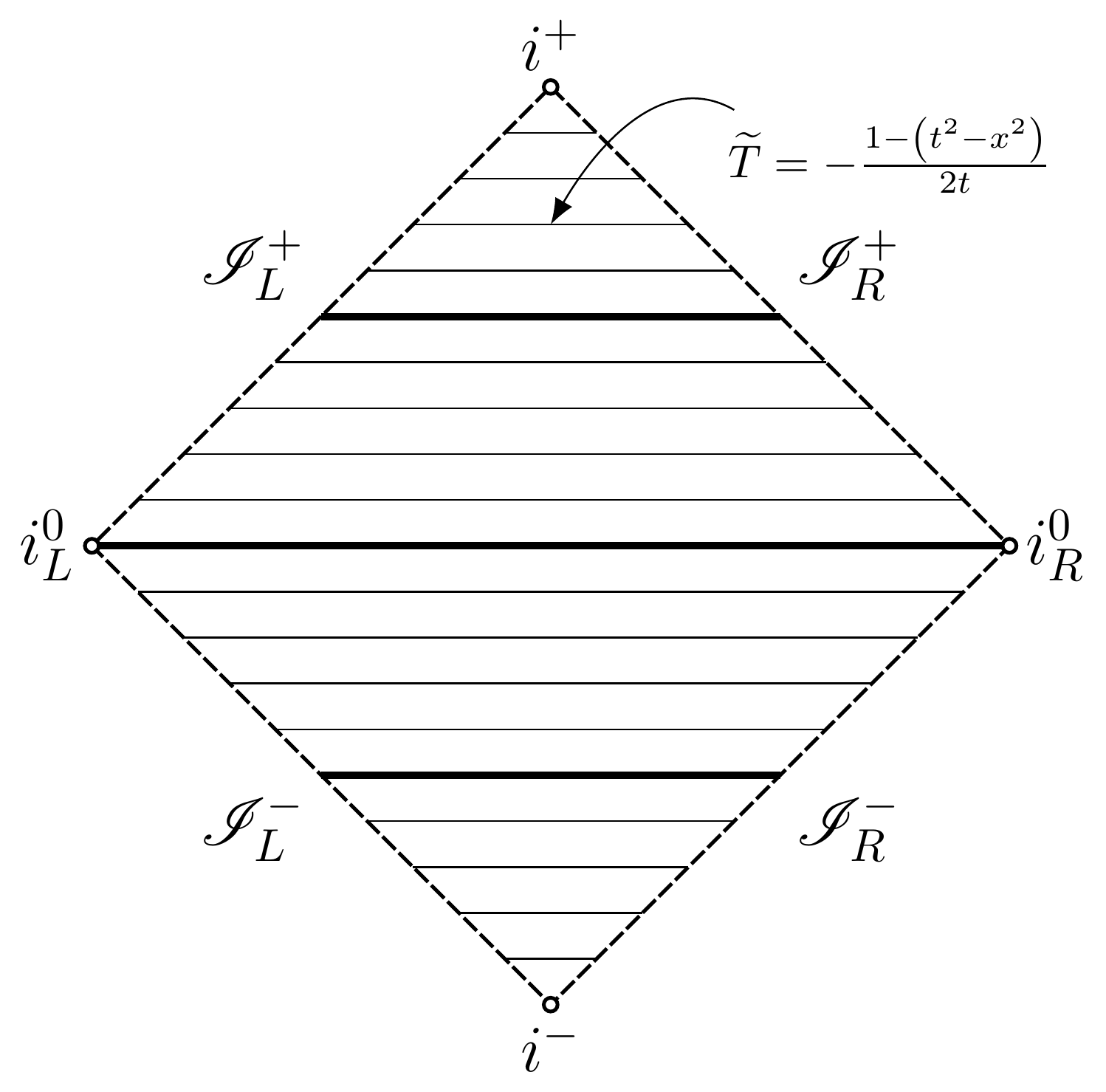}
    \caption{Penrose time slices in standard coordinates on the left and Penrose coordinates on the right. The three limiting surfaces, $T=0$ and $T=\pm\pi/2$, are drawn as thick lines. } 
    \label{fig:penrose}
\end{figure}

To answer this question, we ``decompactify" the time coordinate $T$ via
$$ \tan T = \tan (V+U) = \tan \left(\tan^{-1}(t+x) + \tan^{-1}(t-x)\right) = \frac{2t}{1-(t^2-x^2)}. $$ The Cauchy surface $t=0$ maps to $T=0$. The timelike infinities $t\to\pm\infty$ map to $T=\pm \pi$. The null cone from the origin gives the limiting surfaces with $T=\pm \pi/2$ drawn as thick lines in Fig.~\ref{fig:penrose}. We can get more insight into the nature of the Penrose time function in the domain $|T|\in (0,\pi$) by setting $\tan T\equiv -1/\widetilde{T}$. We get,
$$ t = \widetilde{T} \pm \sqrt{1+x^2+\widetilde{T}^2}, \qquad \mathrm{or} \qquad (t - \widetilde{T})^2 - x^2 = 1+\widetilde{T}^2.$$
This expression is a combination of Milne slicing with time-dependent radii \eqref{eqn:hyperbola}, $t^2 - x^2 = \eta^2$, and time-shifted hyperbolic foliation \eqref{eqn:hyperbolic_foliation}, $(t-\tau)^2 - x^2= 1$.The hyperbolas with time-dependent radii $\sqrt{1+\widetilde{T}^2}$ are shifted along the standard time $t$ by $\widetilde{T}$ (see Fig.~\ref{fig:penrose}). The hyperboloidal nature of the level sets of $T$ is obvious from the Penrose diagram. The straight coordinate lines of the diagram intersect null infinity horizontally and are spacelike globally. We conclude that Penrose coordinates arise from the direct compactification of hyperboloidal coordinates.

\section{Conclusions}\label{sec:conclusions}
The main thesis of this paper is that hyperboloidal coordinates are as natural in Lorentzian manifolds as spherical coordinates are in Riemannian manifolds. Students of general relativity are typically familiar with spherical coordinates and their basic geometric properties. The analogy of hyperboloids to spheres is useful as a teaching tool to introduce counterintuitive features of Lorentzian manifolds related to the unusual metric signature.

Hyperboloidal time functions provide the idealization of surfaces of simultaneity for far-away observers. They maximize spatial volume for a given spacetime volume and have significant advantages in the analysis of wave equations. Unlike standard and Milne time slicings, time-shifted hyperboloidal foliations offer a regular global framework. This regularity allows for an accurate depiction of energy decay, crucial for understanding astrophysical processes and gravitational wave propagation. The central role that hyperboloidal coordinates play in the study of the global causal structure of spacetimes is demonstrated also through the Penrose diagrams drawn in compactified hyperboloidal coordinates. The hyperbolic geometry of hyperboloidal surfaces of simultaneity suggests many interesting questions from holography to computational physics that should be a fruitful research topic in the coming decades.

\section*{Acknowledgements}
I thank members of the Hyperboloidal Research Network for many fruitful interactions. I especially thank Alex Vañó-Viñuales for suggestions on the manuscript and Robert Beig for discussions. This material is supported by the National Science Foundation under Grant No.~2309084.

\section*{CONFLICT OF INTEREST}

The author has no conflicts to disclose.

\bibliographystyle{unsrt}
\bibliography{refs}

\end{document}